\documentclass[10pt,prl,aps,twocolumn,showpacs,floatfix]{revtex4}
\usepackage{graphicx}
\usepackage{amsmath}
\usepackage{amssymb}
\usepackage{bm} 
\usepackage{color}
\usepackage{braket} 

\begin{document}
\draft
\title{Prediction of ultra-narrow Higgs resonance in magnon Bose-condensates}
\author{H. D. Scammell and O. P. Sushkov }
\affiliation{School of Physics, The University of New South Wales,
  Sydney, NSW 2052, Australia}
\date{\today}
\begin{abstract}
Higgs resonance modes in condensed matter systems are generally broad; meaning large decay widths or short relaxation times. This common feature has obscured and limited their observation to a select few systems. Contrary to this, the present work predicts that Higgs resonances in magnetic field induced, three-dimensional magnon Bose-condensates have vanishingly small decay widths.  Specifically for parameters relating to TlCuCl$_3$, we find an energy ($\Delta_H$) to width ($\Gamma_H$) ratio $\Delta_H/\Gamma_H\sim500$, making this the narrowest predicted Higgs mode in a condensed matter system, some two orders of magnitude `narrower' than the sharpest condensed matter Higgs observed so far.  


\end{abstract}
\pacs{64.70.Tg
, 75.40.Gb
, 75.10.Jm
, 74.20.De
}

\maketitle

The Higgs mechanism, and associated Higgs modes, play a central role in modern physics.
The mechanism is responsible for the mass generation of all observed particles in nature, and is the only known, universal mechanism to do so.
Higgs modes are a generic property of systems with a spontaneously broken continuous symmetry. This includes prominent condensed matter phenomena;
superconductivity, Bose condensation (BEC) and superfluidity, quantum magnetism, etc., as well as the Electroweak vacuum. 
Due to the ubiquity and importance of Higgs modes across many branches of physics, their detection has been an exciting, yet difficult, challenge. 
Notably, the discovery of the Electroweak Higgs boson \cite{HiggsCMS, HiggsATLAS} meets both of these descriptions. 
Also attracting a great deal of attention, and proving to host their own difficulties, are the Higgs modes of condensed matter systems. They have been observed in the following settings; 
the charge density wave superconductor 
NbSe$_2$ (1981) \cite{Littlewood1981, Littlewood1982}, three dimensional quantum antiferromagnet (AFM)  
TlCuCl$_3$ (2008) \cite{Ruegg2008}, 
superfluid $^{87}$Rb atoms in an optical lattice (2012) \cite{Endres2012}, 
superconducting NbN (2013) \cite{Matsunaga2013, Matsunaga2014}, and two dimensional quantum 
AFM Ca$_2$RuO$_4$ (2017) \cite{Jain2017}. 
Each setting offers unique insights into the dynamics of Higgs modes and, in particular, the role played by symmetry, dimensionality, as well as the coupling to different degrees of freedom. 
Such factors are seen to have a dramatic influence on the dynamical properties and, ultimately, the observability of the Higgs modes.

A dimensionless parameter characterising the {\it quality} of the Higgs mode is the ratio of the mode
energy over the decay width, $Q= \Delta_H/\Gamma_H$. 
For many interesting, symmetry-broken systems, the Higgs modes obtain $Q \sim 1$, implying such poor quality that the mode
is unobservable. This is the case for the Higgs partners to the following low energy modes: the Higgs partner to spin waves in simple Heisenberg 
AFM; the partner to $\pi$-mesons (chiral symmetry breaking)~\cite{ Albaladejo2012};
the partners to sound in atomic BEC and in superfluid Helium, etc.
It is worth noting that in some cases it is possible to detect Higgs modes indirectly, even for low $Q\sim 1$.
For example, in the case of superconductors NbSe$_2$ and NbN, observation requires either the presence of charge 
density wave order, or the implementation of out-of-equilibrium 
spectroscopy \cite{Cea2014, Cea2015, Krull2016}.
To the best of our knowledge, the narrowest Higgs mode observed in a condensed matter system,
to date, has quality factor $Q=5$. This is the  Higgs resonance in AFM phase of quantum antiferromagnet TlCuCl$_3$ \cite{Ruegg2008}.
The {\it high} quality factor is, in part, due to proximity to the quantum critical point (QCP) \cite{Kulik2011}.

On the other hand, the quality factor of the fundamental $125$ GeV Higgs boson in particle physics is
$Q\approx 2\times 10^4$ ~\cite{Barger2012}. Can one have something comparable in a condensed matter system?
In the present work we argue that it is both possible, and accessible within current experimental techniques. We predict a very narrow Higgs
resonance in BEC of magnons in an external magnetic field. 
The predicted resonance width is so narrow that it may be beyond the resolution of inelastic neutron scattering techniques, with meV resolution \cite{Ruegg2008}. Instead measurement would require $\mu$eV resolution, for which neutron spin-echo technique is appropriate \cite{Bayrakci2006}. Moreover,  Raman spectroscopy, which probes the scalar response channel,  has been used to study Higgs modes of magnon-Bose condensates \cite{Kuroe2008, Kuroe2009}. Raman spectroscopy may therefore provide a suitable means to study the Higgs decay width \cite{com3}.

Although magnon BEC have attracted immense experimental \cite{Ruegg2003, Jaime2004, Kofu2009, Wang2016, Okada2016, Grundmann2016, Kurita2016, Brambleby2017}  and theoretical \cite{Giamarchi1999, Nikuni2000,  Matsumoto2002, Matsumoto2004, Utesov2014, ScammellCritical2017} interest over the past two decades, see reviews \cite{Giamarchi2008, Zapf2014}, the issue of the Higgs magnon width in the BEC phase has not been addressed. Theoretically, the width in the usual AFM phase ({\it i.e.} not the BEC)
was considered in Ref. \cite{Kulik2011} and also in recent Monte Carlo simulations 
\cite{Qin2017,Lohofer2017}.

\begin{figure}[t]
 {\includegraphics[width=0.20\textwidth,clip]{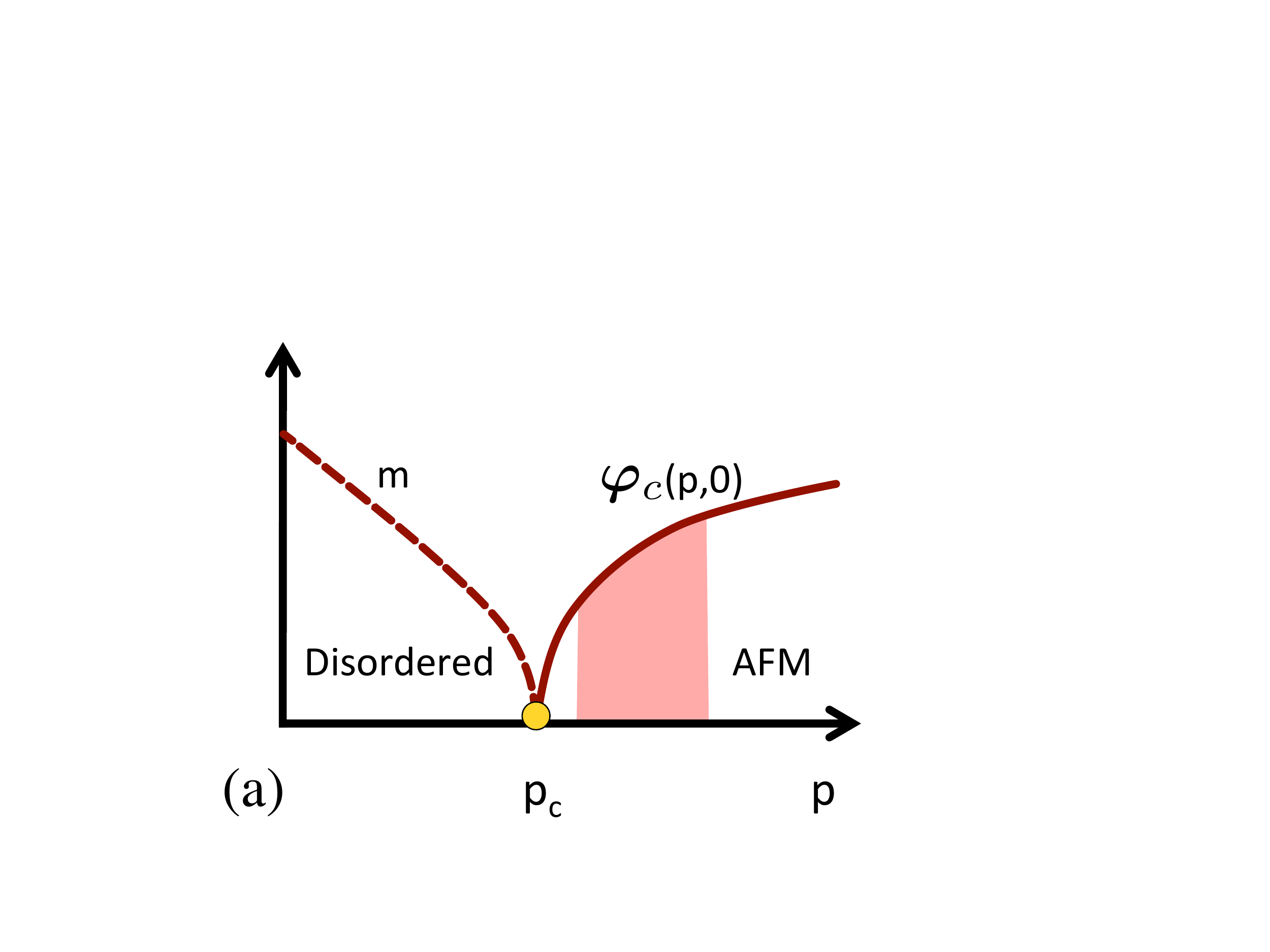} \hspace{0.4cm}}
 { \includegraphics[width=0.241\textwidth,clip]{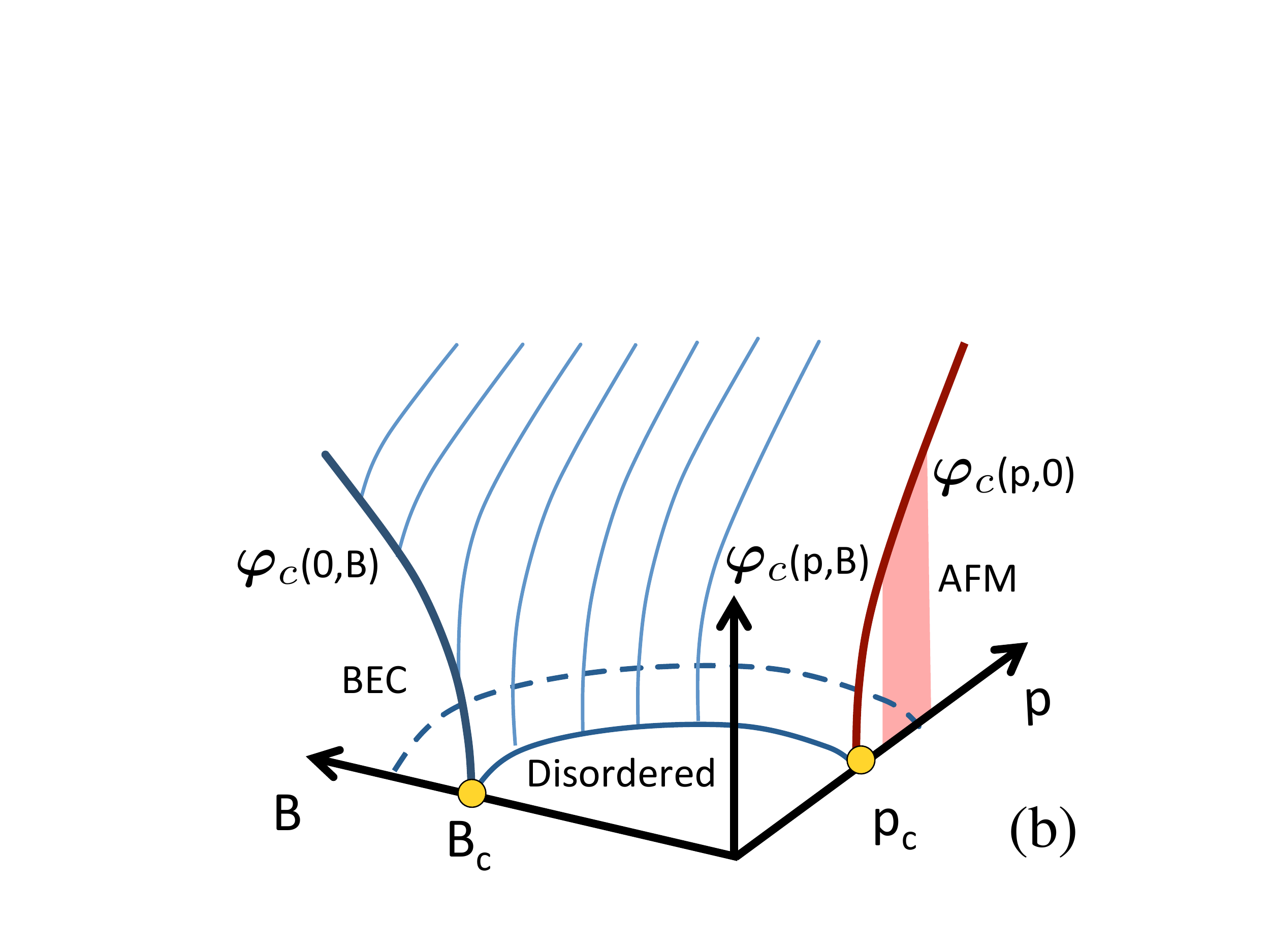}}
\caption{ Zero temperature phase diagram.
Panel a: Magnon gap ($p< p_c$) and spontaneous staggered magnetization ($p> p_c$) 
versus pressure at zero magnetic field, $B=0$.
Panel b: Pressure-Magnetic field phase diagram. 
The vertical axis shows spontaneous staggered magnetization.
The dashed line in the $B-p$ plane shows a contour connecting the simple AFM phase at $B=0$
and the BEC phase at $p=0$.
The red band in both panels indicates the region where the width of the Higgs
excitation has been measured.
 }
\label{PhaseD}
\end{figure}

\begin{figure}[t]
  {\includegraphics[width=0.23\textwidth,clip]{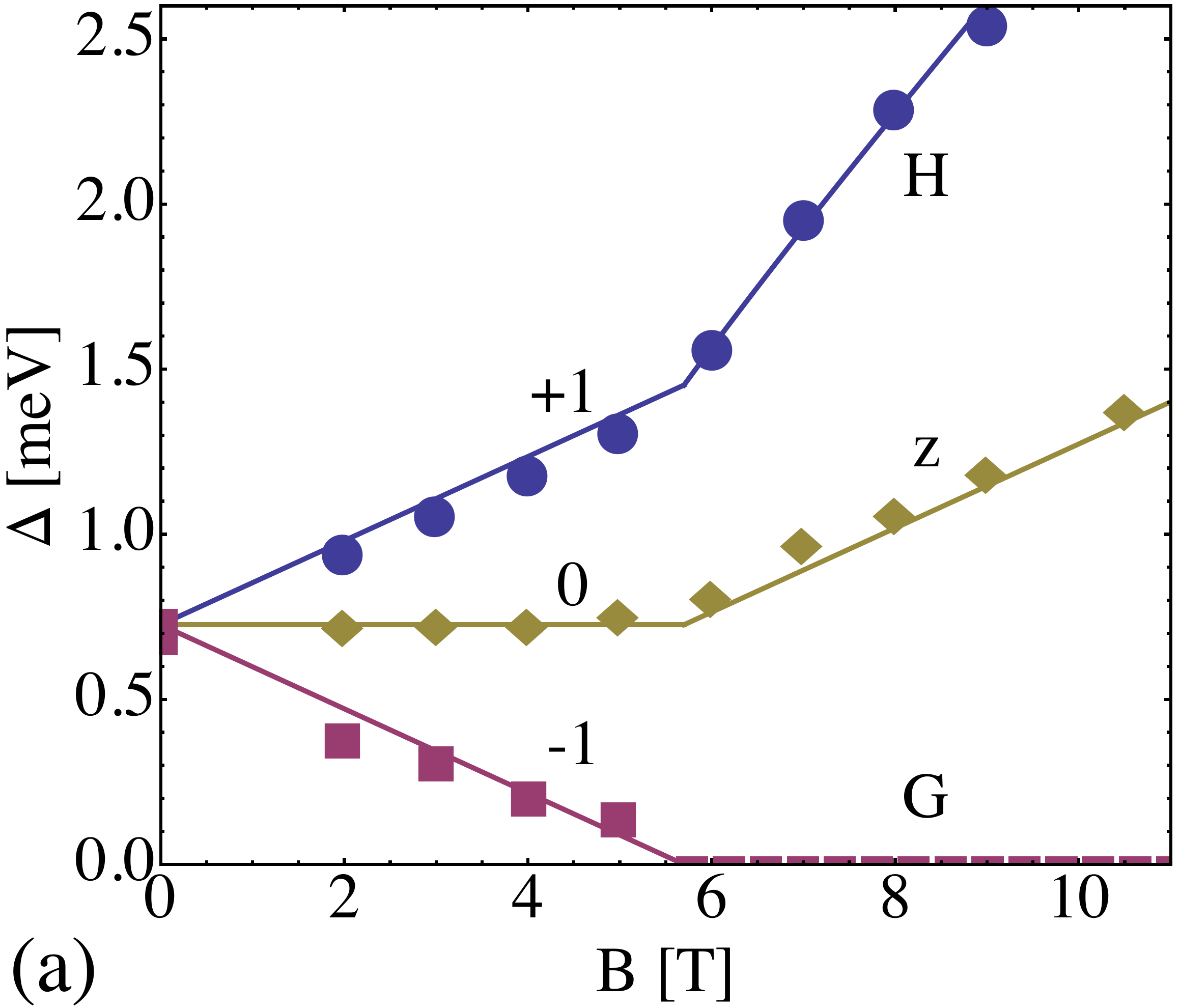}\hspace{0.2cm}}
   {\includegraphics[width=0.23\textwidth,clip]{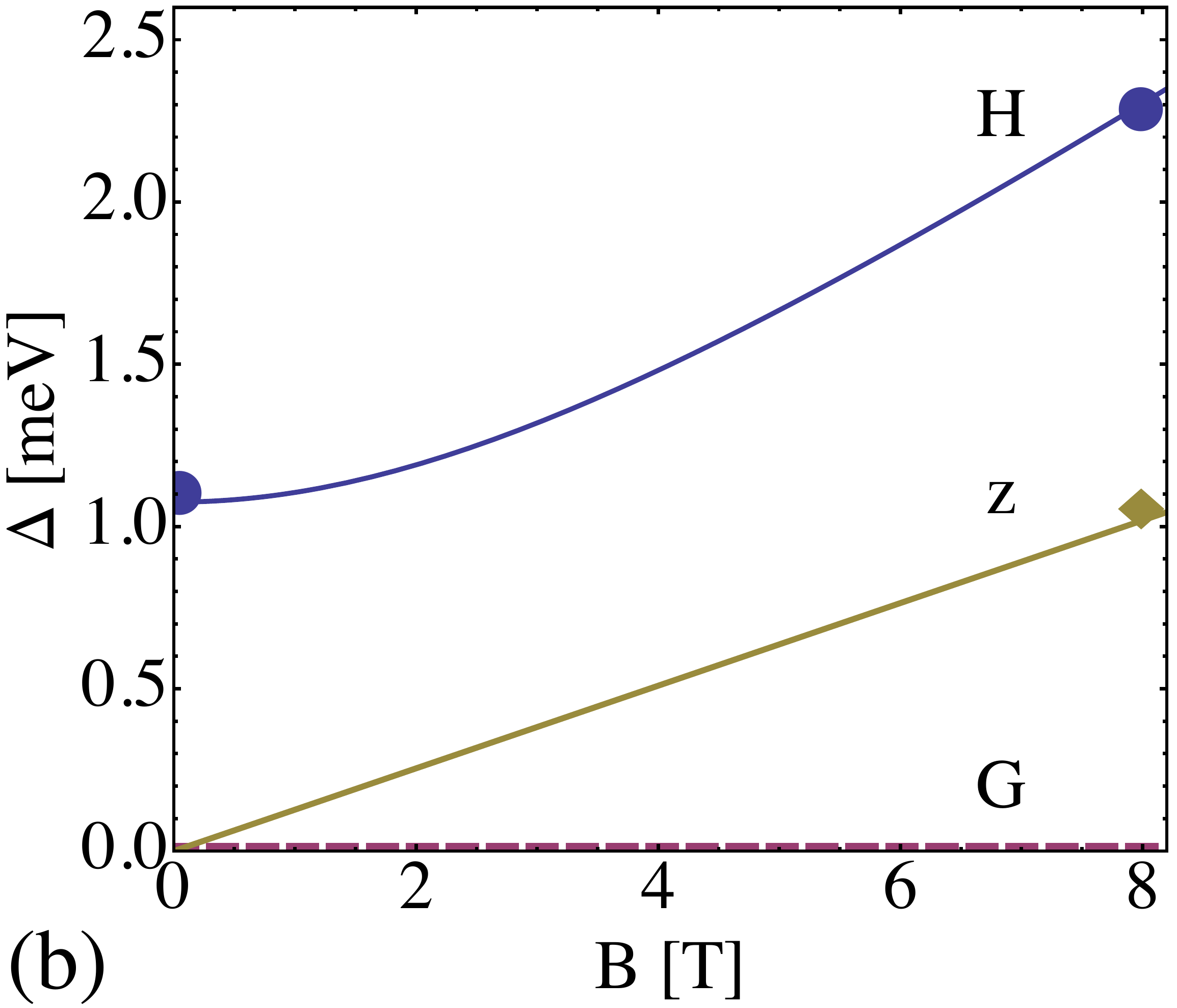}}
\caption{ Evolution of excitation gaps through the BEC phase transition. 
Markers indicate experimental data for TlCuCl$_3$ \cite{Ruegg2002}.
Solid lines are our theoretical results.
Panel a: Gaps versus B-field at zero pressure.
Panel b: Gaps versus B-field along the dashed line in Fig.~\ref{PhaseD}(b)
connecting AFM and BEC phases.
Letters H, z, and G indicate Higgs, z- and Goldstone modes in the ordered phase. While numbers $\{-1,0,+1\}$ refer to Zeeman split modes in the disordered phase, discussed in text.
 }
\label{Disp}
\end{figure}

In this work we address three dimensional (3D) quantum AFMs, having in mind TlCuCl$_3$ and similar.
The zero temperature phase diagram of the system we consider is shown in Fig.~\ref{PhaseD}, where 
Fig. \ref{PhaseD}(a) corresponds to a zero magnetic field slice of Fig. \ref{PhaseD}(b). The quantum phase transition is driven by an 
external parameter, say pressure $p$. At $p > p_c$ the system is in the AFM phase,
the order parameter $\varphi_c\ne 0$ is proportional to the staggered magnetization.
The region $p < p_c$ corresponds to the magnetically disordered phase, where the excitations -- triplons, are gapped and are triply degenerate. We denote by $m$ the gap in the triplon spectrum. 
BEC of magnons at $p < p_c$ can be driven by external magnetic field $B$.
Evolution of the triplon gap under the magnetic field is shown in 
Fig.~\ref{Disp}(a). At weak field there is simple Zeeman splitting of the triple degenerate
gapped triplon. At the critical value of the field $B_c$  the lowest dispersion
branch strikes zero. This is the BEC critical point. At a higher field the lowest
branch remains gapless, this is the Goldstone mode of the magnon BEC. 
Gaps in the middle branch (z-mode) and the top branch (Higgs mode) continue to evolve 
with field. 
The zero temperature $B-p$ phase diagram is presented in Fig.~\ref{PhaseD}(b)
where the vertical axis shows the order parameter. The diagram clearly indicates that 
the AFM phase at $B=0$, $p> p_c$  is continuously connected with  the BEC phase  
at $p=0$, $B > B_c$~\cite{com1}.
Evolution of excitation gaps with magnetic field at zero pressure, $p=0$, is shown in Fig.~\ref{Disp}(a).
Evolution of excitations gaps along the dashed contour in the $B-p$ plane in Fig.~\ref{PhaseD}(b)
is shown in Fig.~\ref{Disp}(b). 
Markers in Fig.~\ref{Disp} indicate experimental data for TlCuCl$_3$ and solid lines represent 
theory described below.

The systems analysed here are close to quantum criticality, where usual 
spin-wave or triplon techniques are insufficient. Instead, the present analysis employs quantum field theory.
The $\sigma$-model-type effective Lagrangian of the system reads~\cite{Sachdev2011,Kulik2011},
\begin{align}   
\label{L1}
{\cal L}[\vec\varphi]&=\frac{1}{2}(\partial_{t}{\vec{\varphi}}-\vec{\varphi}\times\vec{B})^2-\frac{1}{2}(\vec{\nabla}{\vec{\varphi}})^2-\frac{1}{2}m^2{\vec{\varphi}}^{\ 2}-\frac{1}{4}\alpha\vec{\varphi}^{\ 4}.
\end{align}
Here ${\vec {\varphi}}$ is a real vector field describing AFM magnons, $m^2=\gamma^2(p_c-p)$ is the pressure
dependent effective mass ($\gamma$ is a coefficient), and $\alpha$ is the coupling constant.
In Eq. (\ref{L1}) we set the magnetic moment  and the magnon speed equal to unity
$g\mu_B = c=1$. Of course, when comparing with experimental data these quantities have to be restored.
Quantum fluctuations renormalise values of $m^2$ and $\alpha$. The effect of renormalization is 
well understood~\cite{ScammellFreedom2015}; the bare values of $m$ and $\alpha$ are to be replaced
by logarithmically renormalized values, $m \to m_R(\Lambda)$, $\alpha \to \alpha_R(\Lambda)$. 
In our analysis, the renormalization scale, $\Lambda$, is equal to the Higgs gap. We will not discuss logarithmic renormalization any further.
In the spontaneously broken AFM/BEC phase, the classical expectation value immediately follows from (\ref{L1}),
 $\varphi_c^2=(B^2-m^2)/\alpha >0$. Hence the field is
$\vec{\varphi}=(\varphi_c+\sigma,\pi_y,\pi_z)$.
Here we choose $x$-axis to be the direction of the spontaneous magnetization $\varphi_c$ which is orthogonal to the magnetic field
directed along the $z$-axis. Eq. (\ref{L1}) rewritten in terms of dynamic fields, $\sigma$, $\pi_y$, and $\pi_z$,   reads
\begin{eqnarray}
\label{L2}
{\cal L}&=&{\cal L}_2+{\cal L}_3+{\cal L}_4\ ,\\
{\cal L}_2&=&\frac{1}{2}[{\dot \sigma}^2+{\dot \pi}_y^2+{\dot \pi}_z^2]
+B[\sigma{\dot \pi}_y-{\dot \sigma}\pi_y] \ , \nonumber\\
&-&\frac{1}{2}[{\nabla \sigma}^2+{\nabla \pi}_y^2+{\nabla \pi}_z^2]-(B^2-m^2)\sigma^2-\frac{1}{2}B^2\pi_z^2\ ,
\nonumber\\
{\cal L}_3&=& -\alpha\varphi_c\sigma(\sigma^2+\pi_y^2+\pi_z^2)\ ,\nonumber\\
{\cal L}_4&=& -\frac{\alpha}{4}(\sigma^2+\pi_y^2+\pi_z^2)^2\ .\nonumber
\end{eqnarray}
 Performing Fourier transform, $\sigma,\pi \propto e^{-i\omega t+i{\bm k}\cdot{\bm r}}$, 
and using the relation $\frac{1}{2}\varphi^TG^{-1}\varphi= {\cal L}_2$, we obtain the matrix Green's function 
\begin{eqnarray}
\label{G}
G^{-1}=
\left(
\begin{array}{ccc}
-k^2+2(B^2-m^2)&   2iB\omega &   0\\
-2iB\omega & -k^2&  0\\
0 &  0 & -k^2+B^2
\end{array}
\right) \ ,
\end{eqnarray}
where $k=(\omega,{\bm k})$, $k^2=\omega^2-{\bm k}^2$.
Evaluating $|G^{-1}|=0$ results in the following dispersions
for Higgs, Goldstone, and z-modes
\begin{eqnarray}
\label{DM}
&&\omega_{{\bm k}}^{H}=\sqrt{{\bm k}^2+3B^2-m^2+\sqrt{4B^2{\bm k}^2+(3B^2-m^2)^2}}\nonumber\\
&&\omega_{{\bm k}}^{G}=\sqrt{{\bm k}^2+3B^2-m^2-\sqrt{4B^2{\bm k}^2+(3B^2-m^2)^2}}\nonumber\\
&&\omega_{{\bm k}}^{z}=\sqrt{{\bm k}^2+B^2}.
\end{eqnarray} 
Theoretical values of $\omega_{{\bm k}=0}$, given in Eq. (\ref{DM}),
are plotted in Fig.~\ref{Disp} by solid lines~\cite{com2}.
Note that (\ref{DM}) is valid at $B > B_c$ in the spontaneously broken phase. 
In the disordered phase at  $B <  B_c$ there is simple Zeeman splitting of triplon dispersions, 
$\omega_{\bm k}^{(l)}=\sqrt{k^2+m^2}+lB$, where $l=0,\pm 1$.
\begin{figure}[t]
  {\includegraphics[width=0.23\textwidth,clip]{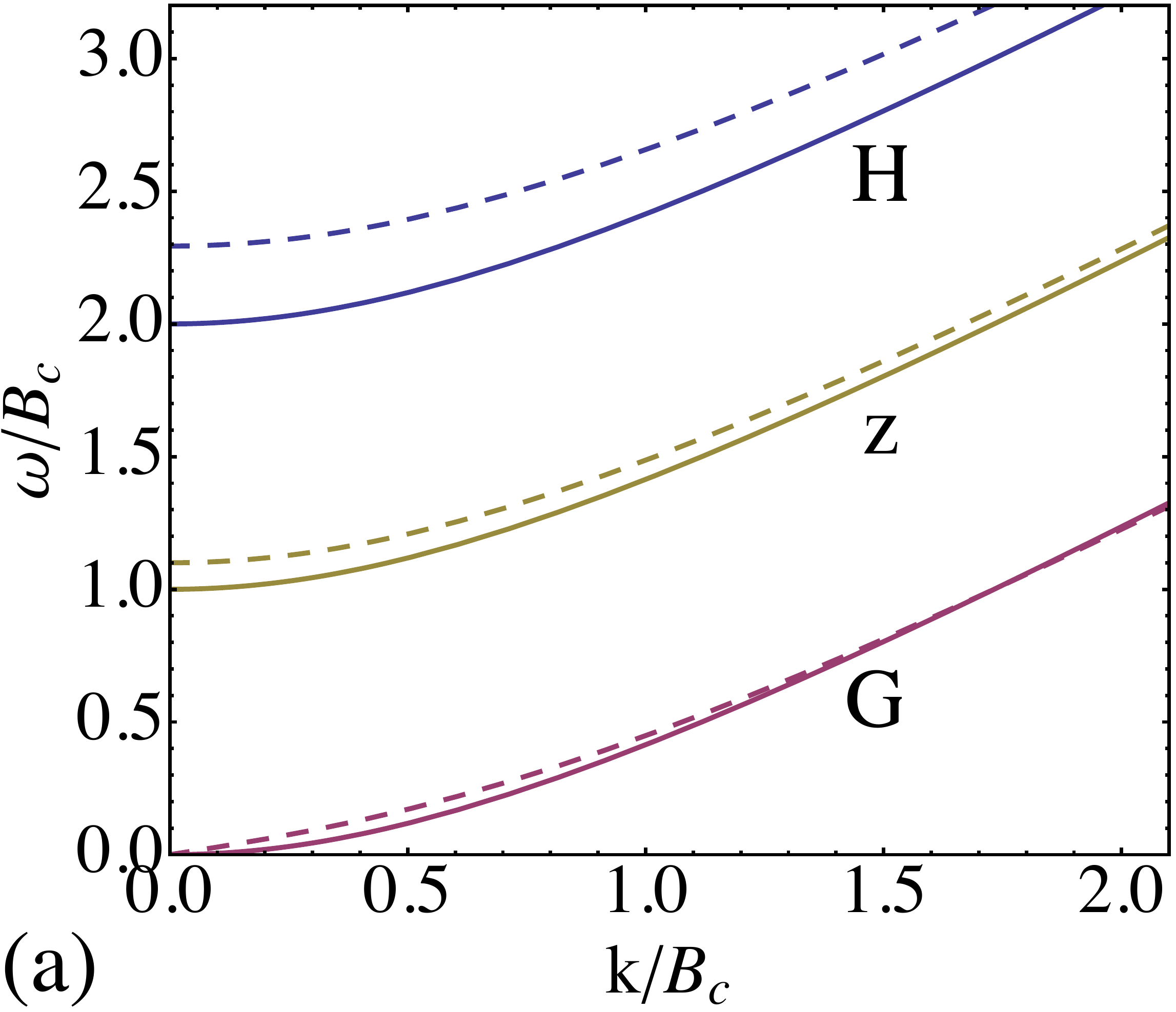}\hspace{0.2cm}}
 {\includegraphics[width=0.23\textwidth,clip]{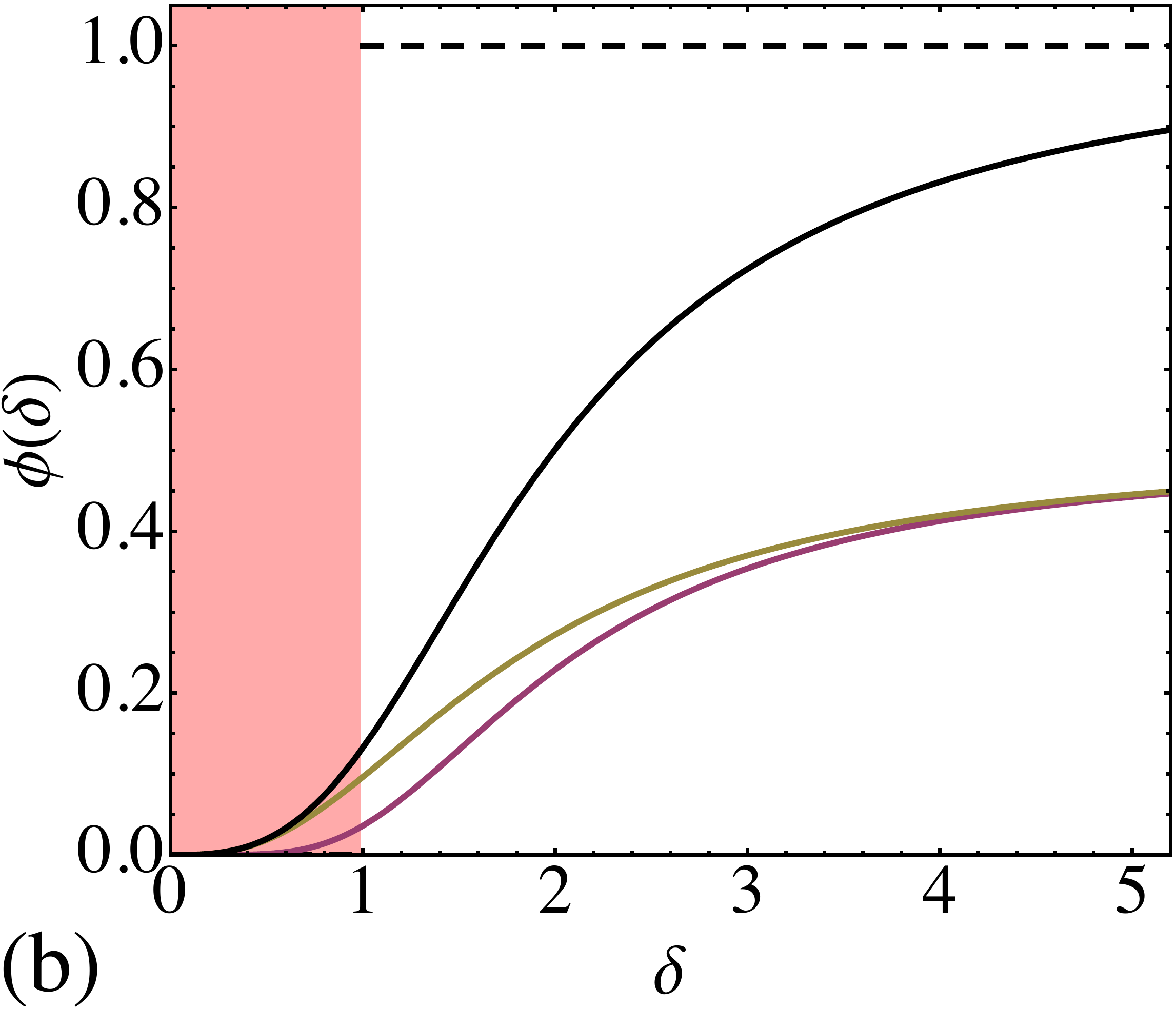}}
\caption{Panel a: Zero spatial momenDispersions of Higgs (H) z- and Goldstone (G) modes for 
$B=B_c$ (solid black) and $B=1.1B_c$ (red dashed).  
Panel b: Scaling functions: $\phi_G(\delta)$ (green) and  $\phi_z (\delta)$ (yellow),  
 Pink shaded region indicates BEC phase, $\delta <1$.
The AFM phase at $B=0$ corresponds to $\delta=\infty$.
 }
\label{Dispk}
\end{figure}
Plots of dispersions (\ref{DM}) versus momentum for two values
of magnetic field, $B=B_c$ and $B=1.1B_c$ are presented in Fig.~\ref{Dispk}(a).
At the critical point, $B=B_c$, the Goldstone mode is quadratic in momentum 
at  $k\to 0$.
At $B>B_c$ the Goldstone mode is linear at $k\to 0$ with a non-linear bend 
at $k \sim B-B_c$.

In principle, the decay amplitude of the Higgs mode is  given by ${\cal L}_3$ and ${\cal L}_4$ in Eq. (\ref{L2}).
However, in the BEC phase there is an important physical and technical complication that
originates from the Berry phase term $B[\sigma{\dot \pi}_y-{\dot \sigma}\pi_y]$ in
${\cal L}_2$, which is not present in the usual AFM phase. 
This term introduces off diagonal kinetic terms into the Lagrangian (or equivalently, into the Green's function (\ref{G})), which cause mixing, or hybridization, between the $\sigma$ and $\pi_y$ modes.
In a standard field theory or in a spin-wave theory without off-diagonal kinetic terms, the field operator is represented
as a combination of corresponding creation and annihilation operators, 
$\sigma \propto \sum_{\bm k} [a_{\bm k} e^{i\omega_{{\bm k}} t-i{\bm k}\cdot{\bm x}}+
a_{\bm k}^\dag e^{-i\omega_{{\bm k}} t+i{\bm k}\cdot{\bm x}}]$. 
This remains valid in the AFM phase of the present work. However, owing to the Berry phase,
this is not valid in the BEC phase. Below we elaborate how to correctly represent the field operators $\sigma$, $\pi_{x/y}$ in terms of creation/annihilation operators, since it is important for understanding the results of the paper.

Let  $a_{\bm k}/a_{\bm k}^\dag$, $b_{\bm k}/b_{\bm k}^\dag$, and $c_{\bm k}/c_{\bm k}^\dag$ 
be the annihilation/creation operators of the Higgs, Goldstone, and z- modes.
Accordingly, the Hamiltonian reads
$H=\sum_{\bm k}\left[\omega_{{\bm k}}^{H}a_{\bm k}^{\dag}a_{\bm k}
+\omega_{{\bm k}}^{G}b_{\bm k}^{\dag}b_{\bm k}
+\omega_{{\bm k}}^{z}c_{\bm k}^{\dag}c_{\bm k}\right]+const$.
It is shown in Section I of the supplementary material that the field operators
are to be expressed in terms of the creation and annihilation operators in the following way
\begin{align}
{\sigma}({\bm x},t)&=\sum_{\bm k} \left\{{\cal A}_{H,{\bm k}}[a_{\bm k} e^{i\omega_{{\bm k}}^H t-i{\bm k}\cdot{\bm x}}+a_{\bm k}^\dag e^{-i\omega_{{\bm k}}^H t+i{\bm k}\cdot{\bm x}}]\right.\nonumber\\
&\left. \hspace{0.9cm}\notag +{\cal A}_{G,{\bm k}}[b_{\bm k} e^{i\omega_{{\bm k}}^G t-i{\bm k}\cdot{\bm x}}+b_{\bm k}^\dag e^{-i\omega_{{\bm k}}^G t+i{\bm k}\cdot{\bm x}}]\right\}\nonumber\\ 
{\pi}_y({\bm x},t)&=\sum_{\bm k} \left\{ {\cal B}_{H,{\bm k}}[a_{\bm k} e^{i\omega_{{\bm k}}^H t-i{\bm k}\cdot{\bm x}}-a_{\bm k}^\dag e^{-i\omega_{{\bm k}}^{\pi} t+i{\bm k}\cdot{\bm x}}]\right.\nonumber\\
&\left. \hspace{0.9cm}\notag + {\cal B}_{G,{\bm k}}[b_{\bm k} e^{i\omega_{{\bm k}}^G t-i{\bm k}\cdot{\bm x}}-b_{\bm k}^\dag e^{-i\omega_{{\bm k}}^G t+i{\bm k}\cdot{\bm x}}]\right\}\nonumber\\
\label{zmode}
\pi_z({\bm x},t)&=\sum_{\bm k} \frac{1}{\sqrt{2\omega_{{\bm k}}^{z}}}
[c_{\bm k} e^{i\omega_{{\bm k}}^{z} t-i{\bm k}\cdot{\bm x}}+c_{\bm k}^\dag e^{-i\omega_{{\bm k}}^{z} t+i{\bm k}\cdot{\bm x}}]
\end{align}
where, denoting $\alpha={H,G}$, the coefficients are
\begin{align}
\notag {\cal A}_{\alpha,{\bm k}}&=\sqrt{
\frac{\omega_{{\bm k}}^{\alpha}B^2}{(B^2+m^2)(\omega_{{\bm k}}^{\alpha})^2+(3B^2-m^2)(2B^2-2m^2+{\bm k}^2)}}\nonumber\\
{\cal B}_{\alpha,{\bm k}}&=D_{\alpha,{\bm k}}{\cal A}_{\alpha,{\bm k}}\ , \ \ \
{\cal D}_{\alpha,{\bm k}}=\frac{-2 i \omega_{{\bm k}}^{\alpha}B}{(\omega_{{\bm k}}^{\alpha})^2-{\bm k}^2} \ .
\end{align}
In a conventional BEC, the Higgs mode is longitudinal and
the Goldstone mode is transverse. As demonstrated in Eqs. (\ref{zmode}), this is
not true for the magnon BEC. Both longitudinal and transverse waves are linear 
combinations of Higgs and Goldstone excitations. 
The bending of the Goldstone dispersion at $k\sim B-B_c$, see Fig.~\ref{Dispk}(a), is a direct manifestation of this hybridisation. 

\begin{figure}[t]
  {\includegraphics[width=0.395\textwidth,clip]{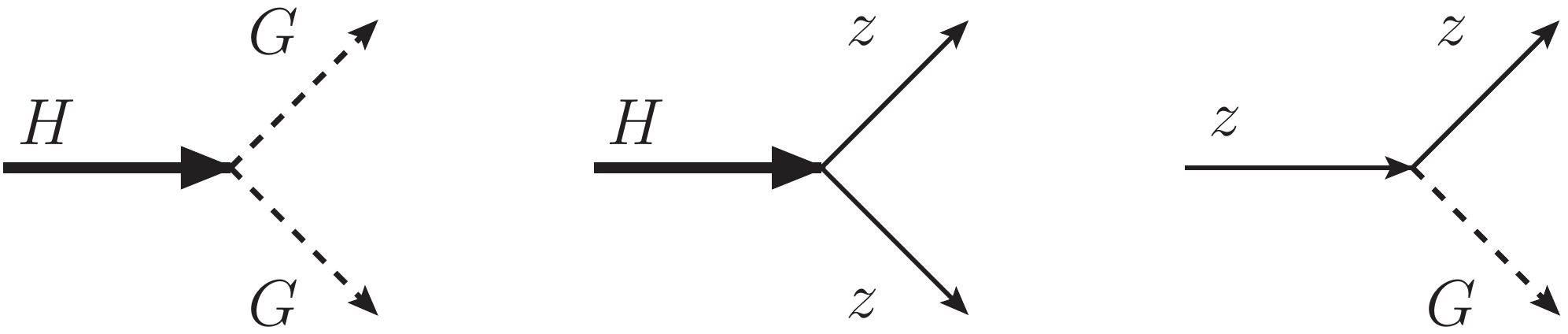}}
\caption{Diagrams for the Higgs and z-mode decay channels. }
\label{Decay}
\end{figure}

From the interaction term ${\cal L}_3$ (\ref{L2}) and Eq. (\ref{zmode}), we conclude that
 Higgs can decay into two Goldstone excitations and into two z-excitations,
as shown diagrammatically in Fig.~\ref{Decay}(a). 
The decay matrix elements follow from Eqs.(\ref{L2}),(\ref{zmode})
\begin{align}
\label{matrixBEC}
\notag{\mathcal M}_{H\to GG}&=
2\alpha\varphi_c {\cal A}_{H,{\bm k}_0}  {\cal A}_{G,{\bm k}_1}  {\cal A}_{G,{\bm k}_2} \times \\ 
\notag&\left\{3+ {\cal D}_{G,{\bm k}_1}{\cal D}_{G,{\bm k}_2}-{\cal D}_{G,{\bm k}_1}{\cal D}_{H,{\bm k}_0}-{\cal D}_{G,{\bm k}_2}
{\cal D}_{H,{\bm k}_0}\right\}\\
{\mathcal M}_{H\to zz}&=2\alpha\varphi_c
\frac{{\cal A}_{H,{\bm k}_0}}{\sqrt{2\omega_{{\bm k}_1}^{z} 2\omega_{{\bm k}_2}^{z}}}  \ .
\end{align}
Here we denote the momentum of initial Higgs by ${\bm k}_0$ and momenta
of final particles by ${\bm k}_1$ and ${\bm k}_2$.
The decay width is given by Fermi golden rule
\begin{eqnarray}
\label{gr}
\Gamma&=&\frac{1}{2}(2\pi)^4\int\frac{d^3k_1}{(2\pi)^3}\frac{d^3k_2}{(2\pi)^3}
\nonumber\\
&\times&
|{\mathcal M}|^2\delta(\omega_{{\bf k}_0}-\omega_{{\bf k}_1}-\omega_{{\bf k}_2})
\delta({\bm k}_0-{\bm k}_1-{\bm k}_2),
\end{eqnarray}
where the coefficient $1/2$ stands to avoid double counting of final bosonic states.
A direct integration gives the decay widths. For clarity, and to 
avoid lengthy formulas, we present here only the partial widths in rest frame,
 ${\bm k}_0=0$, $\omega^H=\sqrt{2(3B^2-m^2)}$,
\begin{align}
\label{pZ}
\frac{\Gamma_{H\to GG}}{\omega^H}&=\frac{\alpha_0}{8\pi}\phi_G(\delta)\\
\notag\frac{\Gamma_{H\to zz}}{\omega_H}&=\frac{\alpha_0}{8\pi}\phi_z(\delta) \ ,
\end{align}
where the scaling functions $\phi_G$ and $\phi_z$ depend on the parameter,
$\delta^2=\alpha\varphi_c^2/B^2=1-m^2/B^2$
\begin{eqnarray}
&&\phi_G(\delta)=\frac{\delta^2}{2}
\left(3-\frac{2(2+\delta^2)-4(\delta^2-\sqrt{\delta^4+2\delta^2+4})}{(\delta^2-\sqrt{\delta^4+2\delta^2+4})^2}\right)^2\nonumber\\
&&\times\frac{(2+\sqrt{\delta^4+2\delta^2+4})\sqrt{2-\delta^2+2\sqrt{\delta^4+2\delta^2+4}}}
{(2+\delta^2)^{3/2}
\sqrt{\delta^4+2\delta^2+4}(2+\delta^2+\sqrt{\delta^4+2\delta^2+4})^2}\ , \nonumber\\
&&\phi_z(\delta)=\frac{\delta^3/2}{(2+\delta^2)^{3/2}}\ .
\end{eqnarray}
The parameter, $\delta$, ranges from $\delta=0$ at a BEC critical point at; $p < p_c$,
$m^2 > 0$, and $B=m$, to $\delta = \infty$ in the AFM phase; $p > p_c$,
$m^2 < 0$, and $B=0$. 
Asymptotics of the scaling functions are: 
$\phi_G(\infty)=\phi_z(\infty)=1/2$, $\phi_G(\delta \to 0)\propto \delta^6$,
$\phi_z(\delta\to 0) \propto \delta^3$.
Plots of $\phi_G(\delta)$ and $\phi_z(\delta)$ are presented in
Fig.~\ref{Dispk}(b). A very strong suppression at small $\delta$ is evident, and this constitutes our main physical result.

In the AFM phase at $B=0$, $\delta=\infty$, the total width 
$\Gamma=\Gamma_{H\to GG}+\Gamma_{H\to zz}$
has been already measured~\cite{Ruegg2008} and 
calculated~\cite{Kulik2011,Qin2017,Lohofer2017}. 
In this regime, Eq. (\ref{pZ}) gives $\Gamma/\omega^H=\alpha/8\pi$
which is consistent with previous work~\cite{Ruegg2008,Kulik2011,Qin2017,Lohofer2017}.
The most interesting  prediction of Eq. (\ref{pZ}) is the dramatic suppression of the
width at $\delta < \infty$ and especially in the BEC phase ($\delta < 1$).
Explicitly for TlCuCl$_3$, taking $p=0$ and $B=1.1B_c\approx 6.4$T, the Higgs gap
is $\omega^H\approx 1.6$meV and the quality factor $Q= \omega^H/\Gamma \approx 500$, predicted by Eq. (\ref{pZ}),
is much higher than that in the AFM phase ($Q\approx 5$) which
itself is the highest Higgs quality factor observed in a condensed matter system
so far. Even at the BEC border line, $p=p_c$, $\delta=0$, the quality factor is $Q\approx50$.

It is interesting to note that the z-mode is even more narrow than the Higgs. The dominant decay channel is via emission of the
Goldstone excitation, $z\to z+G$,  Fig.~\ref{Decay}(b) (double Goldstone emission is also possible, $z\to z+G+G$, but with lower amplitude). Moreover, the decay is possible only if the speed of $z$ 
is higher than the speed of Goldstone excitation, making this a magnetic analog of Cherenkov 
radiation.
Due to ${\cal L}_4$ in Eq. (\ref{L2}) the Higgs mode can also decay into three Goldstones,
$H\to G+G+G$.
However, the probability of this decay mode is much smaller than $H \to G+G$  considered above due to a reduced phase space and being at next-order in perturbative coupling $\alpha$.
At non-zero temperature Raman processes become possible, $H+G \to H+G$, $z+G \to z+G$. 
The corresponding broadening can be calculated using the developed technique supplemented with appropriate Bose occupation factors. However, at
low $T$, the Raman broadening is small due to Bose occupation factors, and therefore we do not consider it here.
Comparing with real compounds one often has to account for weak spin-orbit anisotropy; we explicitly treat this scenario in Section II of the supplementary material.
The anisotropy slightly changes one or more mode dispersion. Which mode is affected depends on the orientation of the magnetic field. If magnetic field is oriented such that the anisotropy only shows up as an additional gapping of the z-mode, the Higgs partial decay into the z-mode will be further reduced due to phase space, strengthening the present conclusions.
Finally, it is worth noting that in the BEC phase the width becomes 
so narrow that the decay into two phonons may be comparable with purely magnetic decay
mechanisms considered here.

In conclusion, we predict that Higgs modes in a magnon Bose-condensate phase of 3D
quantum magnets are ultra-narrow, {\it i.e.} have vanishingly small decay width. We demonstrate that the Higgs mode in the Bose-condensate phase can be tuned to have a decay width two orders of magnitude smaller than the corresponding mode in the antiferromagnetic phase, which itself is the narrowest Higgs mode observed so far in a condensed matter system. An essential feature of the Bose-condensate is a Berry phase contribution that causes the collective modes; Higgs and Goldstone, to appear as the hybridisation of longitudinal and transverse excitations of the condensate order parameter. This hybridisation plays a key role in {\it narrowing} the Higgs mode. Moreover, we calculate dispersions of all collective excitations in the 
magnon Bose-condensate phase and find that hybridisation also manifests itself as a bending of the dispersion branches, thus providing further experimental tests of the scenario posed here.


\begin{thebibliography}{43}
\bibitem[{com({\natexlab{b}})}]{com1}
\bibinfo{note}{Considering the AFM part of the phase diagram Fig.
  \ref{PhaseD}(b) one has to remember that an arbitrary weak magnetic field
  orients the AFM macroscopic polarization perpendicular to the field.}
\end{thebibliography}
\end{document}